\documentclass{article}

\usepackage{cite}
\usepackage{hyperref}
\usepackage{graphicx}
\usepackage{amsmath}
\usepackage[utf8]{inputenc}

\title{DISCOVER: A Physics-Informed, GPU-Accelerated Symbolic Regression Framework}

\author{
Udaykumar Gajera$^{1,2}$,
Mohsen Sotoudeh$^{3,4}$,
Kanchan Sarkar$^{2,3}$,
Axel Gro{\ss}$^{2,3}$
}

\date{}

\begin{document}
\maketitle

\noindent
$^1$ Department of Physics and CSMB, Humboldt-Universit{\"a}t zu Berlin, Berlin, Germany\\
$^2$ Institute of Theoretical Chemistry, Ulm University, Oberberghof 7, 89081 Ulm, Germany\\
$^3$ Helmholtz Institute Ulm (HIU), Electrochemical Energy Storage, 89081 Ulm, Germany\\
$^4$ Karlsruhe Institute of Technology (KIT), P.O. Box 3640, D-76021 Karlsruhe, Germany

\section*{Summary}

Symbolic Regression (SR) enables the discovery of interpretable mathematical relationships from experimental and simulation data. These relationships are often coined descriptors which are defined as a fundamental materials property that is directly correlated to a desired or undesired functional property of the material. Although established approaches such as Sure Independence Screening and Sparsifying Operator (SISSO) have successfully identified low-dimensional descriptors within large feature spaces \cite{ouyang2018sisso}, many existing SR tools integrate poorly with modern Python workflows, offer limited control over the symbolic search space, or struggle with the computational demands of large-scale studies. This paper introduces \textsc{DISCOVER} (Data-Informed Symbolic Combination of Operators for Variable Equation Regression), an open-source symbolic regression package developed to address these challenges through a modular, physics-motivated design. \textsc{DISCOVER} allows users to guide the symbolic search using domain knowledge, constrain the feature space explicitly, and take advantage of optional GPU acceleration to improve computational efficiency in data-intensive workflows, enabling reproducible and scalable SR workflows. The software is intended for applications in computational physics, computational chemistry, and materials science, where interpretability, physical consistency, and execution time are especially important, and it complements general-purpose SR frameworks by emphasizing the discovery of physically meaningful models \cite{wang2025symantic}.

\section*{Statement of Need}

Symbolic regression is widely used in scientific domains where interpretability and physical insight are essential, including physics, chemistry, and materials science. This insight can be expressed as a descriptor which corresponds to a correlation between a fundamental materials property and a desired or desired  function of the material \cite{Sotoudeh2022} While many SR methods can recover analytical expressions from data \cite{udrescu2020aifeynman}, practical adoption is often limited by several factors: insufficient integration with Python-based scientific workflows, limited mechanisms for incorporating \textit{a priori} physical knowledge, and high computational cost when exploring large symbolic search spaces. These challenges make it difficult for researchers to apply SR methods efficiently and reproducibly in real-world scientific studies.

Existing tools such as SISSO provide powerful, deterministic strategies for identifying sparse descriptors but are not designed to offer fine-grained, user-defined control over the symbolic search or to leverage modern hardware acceleration as a core feature \cite{ouyang2018sisso, purcell2023recentadvancessissomethod}. Conversely, more flexible or physics-informed SR approaches (PiSR) may require complex customization or lack scalable performance \cite{sun2022physics}. As a result, researchers often face trade-offs between interpretability, usability, and computational efficiency.

Recent symbolic regression tools have demonstrated impressive capabilities in recovering analytical expressions from data. For example, AI Feynman \cite{udrescu2020aifeynman} leverages symbolic manipulation and neural-guided search to rediscover known physical laws, while extensions of the SISSO framework, such as SISSO++ \cite{purcell2023recentadvancessissomethod}, continue to advance large-scale descriptor discovery through efficient sparsity-driven screening. These methods represent important progress in the field; however, they often prioritize either fully automated discovery or highly specialized workflows, and may offer limited flexibility for incorporating fine-grained physical constraints, modern Python integration, or hardware acceleration as first-class features.

\textsc{DISCOVER} addresses this gap by providing a Python-native symbolic regression framework that explicitly supports physics-informed constraints and optional GPU-accelerated computation. By allowing users to define constraints on operators, feature combinations, and physical consistency through a configuration-based interface, \textsc{DISCOVER} lowers the barrier to incorporating domain knowledge into SR workflows. Its design supports reproducible experimentation, efficient exploration of constrained search spaces, and seamless integration into existing scientific Python ecosystems.

\section*{Software Description}

\textsc{DISCOVER} is an open-source symbolic regression package designed for the guided discovery of interpretable mathematical expressions. The software generates candidate symbolic expressions from user-provided features and operator libraries, evaluates them against target data, and identifies parsimonious models that balance accuracy and simplicity. The search process is iterative and incorporates pruning strategies informed by user-defined physical constraints. To support sparse model discovery, \textsc{DISCOVER} provides access to multiple sparsifying search strategies, including heuristic, optimization-based, and stochastic approaches such as Orthogonal Matching Pursuit (OMP) \cite{omp}, Mixed-Integer Quadratic Programming (MIQP) \cite{miqp}, and Simulated Annealing \cite{sa}.

The software architecture is modular and Python-native, enabling straightforward integration with common scientific libraries. Computationally intensive operations such as feature generation and model evaluation are parallelized and executed on hardware accelerators when available. For large-scale studies, \textsc{DISCOVER} supports optional GPU acceleration via CUDA on NVIDIA GPUs and Metal Performance Shaders (MPS) on Apple Silicon devices, while maintaining efficient CPU-based execution for standard workloads. This hardware-aware design enables scalable symbolic regression workflows on both high-performance computing systems and local development environments.

\begin{figure}[h!]
\centering
\includegraphics[width=10cm]{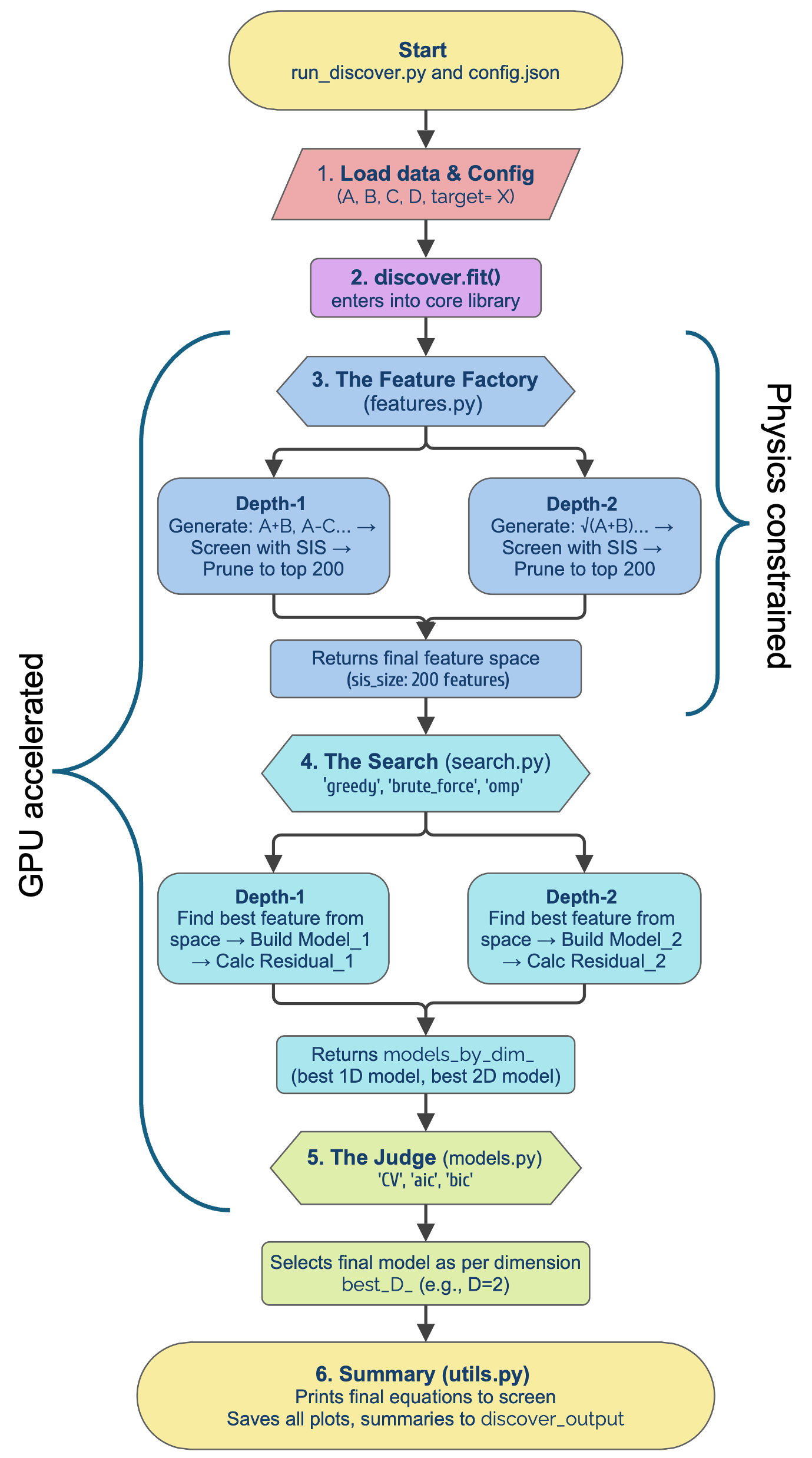}
\caption{Overview of the \textsc{DISCOVER} workflow, illustrating iterative feature generation, physics-informed screening, and sparse model selection.}
\label{fig:workflow}
\end{figure}

\subsection*{Core Optimization Objective}

All search strategies implemented in \textsc{DISCOVER} are designed to approximate or solve a common underlying optimization problem. Given a set of $M$ candidate symbolic features $\boldsymbol{\Phi} = \{\Phi_1, \Phi_2, \ldots, \Phi_M\}$ and a target property vector $\mathbf{y}$, the objective is to identify a sparse linear combination of features that accurately models the data. This problem can be expressed as an $L_0$-regularized least-squares regression:

\begin{equation}
\min_{\boldsymbol{\beta}} \; \left\| \mathbf{y} - \boldsymbol{\Phi} \boldsymbol{\beta} \right\|_2^2
\quad \text{subject to} \quad
\left\| \boldsymbol{\beta} \right\|_0 \le D,
\end{equation}

where $\boldsymbol{\beta}$ is the coefficient vector and $\|\boldsymbol{\beta}\|_0$ denotes the number of nonzero entries, enforcing a maximum descriptor dimensionality $D$. This formulation is common in sparse symbolic regression and descriptor discovery and is known to be NP-hard. To put it simply, there is no known mathematical shortcut to efficiently find the exact optimal solution for this type of problem. Therefore, instead of trying to calculate the impossible 'perfect' answer, we must rely on smart approximation strategies to find a high-quality model within a reasonable timeframe. As a result, \textsc{DISCOVER} offers multiple heuristic, approximate, and specialized search strategies to explore this objective efficiently under user-defined physical and computational constraints.

\subsection*{Physics-Informed Constraints}

A central design goal of \textsc{DISCOVER} is to facilitate the explicit incorporation of domain knowledge into the symbolic regression process. Physical constraints are specified through a configuration-based interface and applied during expression generation and evaluation. Dimensional consistency is enforced through integration with the \texttt{pint} unit-handling library, enabling unit-aware symbolic operations and validation of candidate expressions. By tracking physical units throughout the search process, \textsc{DISCOVER} can exclude dimensionally invalid expressions early, reducing the effective search space and promoting the discovery of physically meaningful and interpretable models.

\section*{Design Philosophy and Constraints}

A core design goal of \textsc{DISCOVER} is to enable direct incorporation of domain expertise into the symbolic regression process. Rather than relying solely on automated sparsity or heuristic search\cite{doi:10.1021/acsami.2c12848}, \textsc{DISCOVER} allows users to specify constraints via a configuration file without modifying source code. Supported constraints include enforcement of dimensional consistency \cite{Tenachi_2023}, restrictions on allowed operators or expression complexity, and user-defined rules governing variable combinations and functional forms \cite{bladek2021shape}.

These constraints reduce the effective search space, improve interpretability, and help ensure that discovered expressions are physically meaningful. This approach is particularly useful in scientific domains where prior knowledge is well established and model plausibility is as important as predictive accuracy \cite{keren2025framework}.

\section*{Scope and Use Cases}

\textsc{DISCOVER} is intended for scientific applications where symbolic regression is used as a tool for model discovery rather than purely predictive performance. Typical use cases include identifying low-dimensional descriptors for physical or chemical properties, such as crystal structure stability \cite{gajera2022toward} or ion mobility in energy storage materials \cite{SOTOUDEH2024101494}. The software is especially suited to computational physics, computational chemistry, and materials science workflows that benefit from Python integration and hardware-accelerated computation, spanning from battery cathode discovery \cite{ziheng49309} to accurate discrimination of magnetic structure \cite{PhysRevB.108.014403}.

\section*{Limitations}

The effectiveness of \textsc{DISCOVER} depends on the quality of the input features and the appropriateness of user-defined constraints. Overly restrictive constraints may exclude valid expressions, while insufficient constraints can lead to large search spaces with increased computational cost. Although GPU acceleration improves performance for many workloads, \textsc{DISCOVER} is not optimized for fully unconstrained searches over extremely large feature spaces compared to specialized low-level implementations such as SISSO \cite{ouyang2018sisso}. Ongoing development focuses on expanded operator libraries, improved benchmarking, and scalability enhancements.

\section*{Statement of AI Assistance}
During the preparation of this work, the authors used large language models to assist in refactoring the source code. Specifically, AI tools were utilized to remove redundant functions, generate explanatory comments for complex logic, and standardize function naming conventions (e.g., renaming legacy short-form functions like \texttt{r\_dis()} to the more descriptive \texttt{run\_discover()}).

\section*{Acknowledgements}

U.G. acknowledges primary support from the NFDI consortium FAIRmat, funded by the Deutsche Forschungsgemeinschaft (DFG) under project 460197019. Furthermore, this work contributes to the research performed at CELEST (Center for Electrochemical Energy Storage Ulm-Karlsruhe). Support by the German Research Foundation (DFG) through the POLiS Cluster of Excellence, Project ID 390874152, and by the Dr. Barbara Mez-Starck-Foundation is gratefully acknowledged.

\bibliographystyle{unsrt}
\bibliography{references}

\end{document}